\documentclass{desyproc}

\begin{document}
\title{Pair-production opacity at high and very-high gamma-ray energies}

\author{{\slshape Dieter Horns$^1$, Manuel Meyer$^{1,2}$}\\[1ex]
$^1$Institut f\"ur Experimentalphysik, Universit\"at Hamburg , Germany\\
$^2$Oskar-Klein Centrum, Stockholm, Sweden}

\contribID{horns\_dieter}

\desyproc{DESY-PROC-2013-XX}
\acronym{Patras 2013} 
\doi  

\maketitle

\begin{abstract}
The propagation of high energy (HE, $E_\gamma>100$~MeV) 
and very high-energy gamma-rays (VHE, $E_\gamma>100$~GeV) 
in the extra-galactic photon field 
leads to pair-production and consequently 
energy- and distance-dependent attenuation of the primary intensity. 
The spectroscopy of an increasing number of extra-galactic objects 
at HE and VHE energies has demonstrated indeed the presence of such 
an attenuation which in turn has been used to constrain the photon 
density in the medium. At large
optical depth ($\tau\gtrsim 2$) potential modifications of
pair-production due to competing but rare processes (as, e.g., 
the presence of sub-neV axion-like particle) may be found. Indications
for a pair-production anomaly have previously been found with VHE-spectra. 
Here, we present further indications (at the level of
$3.68~\sigma$) for a reduced optical depth at high energies from an analysis of Fermi-\textit{LAT} data.
\end{abstract}

\section{Introduction}
\begin{figure}[hb]
\parbox{0.65\linewidth}{
\includegraphics[width=\linewidth]{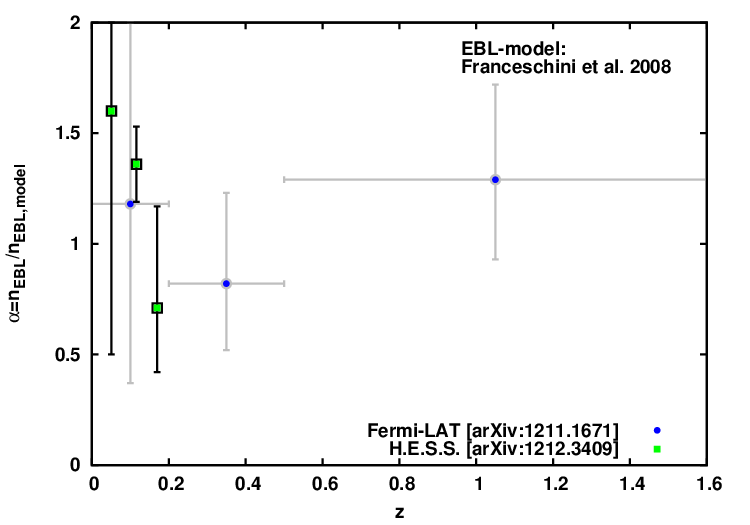}}
\parbox{0.33\linewidth}{
\caption{Comparison of the high-energy ($E_\gamma>100$~MeV) 
and VHE ($E_\gamma>100$~GeV) measurements of the
normalization factor of the extra-galactic background light (EBL)
for different ranges of redshift. }\label{Fig:ebl_level}}
\end{figure}

The extra-galactic photon field in the optical/ultraviolet and infra-red is the
stellar and dust-reprocessed light (see \cite{Dwek:2012nb} for a review)
accumulated during the cosmological evolution following the era of
re-ionization.  For sufficiently energetic ($E> 10$~GeV) photons from distant
sources, pair-production processes with this background photon field 
lead to an energy- and distance dependent exponential attenuation, $\exp(-\tau)$,
where $\tau$ is the optical depth.
This effect has recently been detected in the observed HE gamma-ray spectra of
50 Blazars in the redshift range 0.5 to 1.6 \cite{Ackermann:2012sza}
as well as independently in the observed VHE gamma-ray spectra of mainly
2 BL Lac type objects at redshifts of $0.116$ and $0.186$ \cite{Abramowski:2012ry}. Given the measurement uncertainties, the spectral shape of the 
extra-galactic background light (EBL) has been fixed to a choice of models 
with  normalizations left to vary. The two independent measurements of the 
redshift dependent EBL level for one particular model \cite{Franceschini:2008tp}
is shown in Fig.~\ref{Fig:ebl_level}. 
Given the large uncertainties, variations of the normalization by a factor 
of two are consistent with the data. Particularly, the VHE data favor
a drop of the EBL normalization towards larger redshifts, broadly consistent 
with the HE measurement.\\
At large optical depth ($\tau> 2$), 
modifications of the transparency by non-standard propagation
effects may lead to noticable effects in the attenuation. Even though
in principle the residuals (Fig.~4 of \cite{Abramowski:2012ry} and
Fig.~2 of \cite{Ackermann:2012sza}) do not show obvious deviations from
the best-fit, it is difficult to interpret this result 
given that the normalization
of the EBL (and therefore of the optical depth) was varied by more than a factor of
two between the different redshift bins.\\
Several studies  of the VHE measurements 
provided indications for deviations from the expected transparency 
 \cite{Aharonian:2007wc,Aliu:2008ay,Essey:2011wv,Horns:2012fx}. 
The proposed interpretations have either focussed on the assumption
that Blazars are powerful accelerators of ultra-high energy protons 
\cite{Essey:2009zg} or additional processes including light
pseudo-scalars (a la axion-like particles, for
a review see, e.g., \cite{Jaeckel:2010ni})  have been invoked
\cite{DeAngelis:2007dy,Meyer:2013pny}. In this contribution, we extend our
previous work in the HE energy regime (see also \cite{Meyer:2012sb} for an update of the original analyses of VHE data \cite{Horns:2012fx})
using Fermi-\textit{LAT}
observations of distant AGN \cite{Meyer:2013}.\footnote{Note, the search for anomalous transparency
effects as discussed here is complementary to searches for additional 
noise induced by photon-axion coupling in AGN spectra as originally proposed 
for optical QSO spectra \cite{Ostman:2004eh}
and recently extended to a VHE and X-ray spectra \cite{wouters:2013sc}.}

\section{Fermi-\textit{LAT} observations}
\begin{figure}[hb]
\parbox{0.65\linewidth}{
\includegraphics[width=\linewidth]{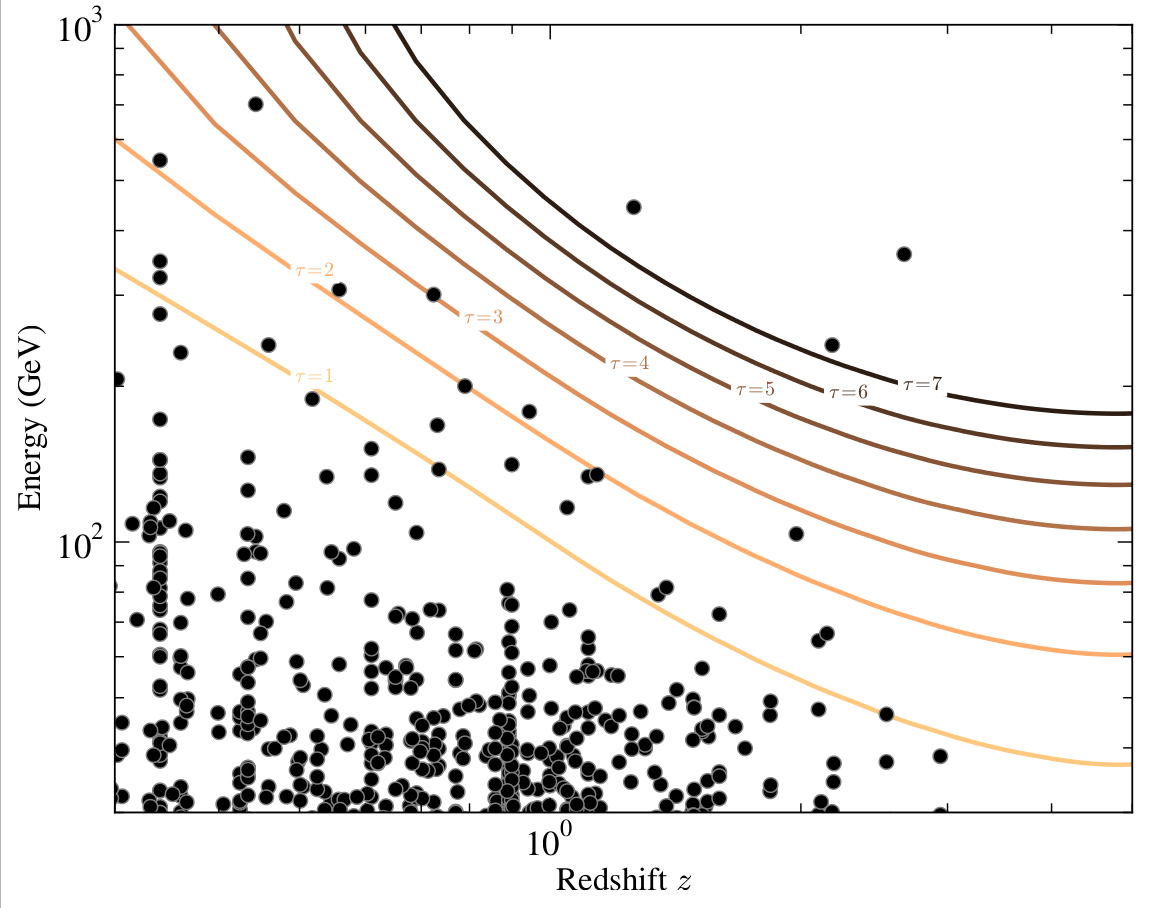}
}
\parbox{0.33\linewidth}{
\caption{Photons detected at energies $E>10$~GeV associated with 
$\gamma$-ray emitting AGN with known redshift.\label{Fig:photons}}
}
\end{figure}

The data-set from the first 4.3 years of operation (until Nov. 29, 2012)\footnote{using time-intervals passing
the standard cuts} of Fermi-\textit{LAT} are
searched for the most energetic photons which can be associated with known
gamma-ray emitting AGN from the second Fermi-\textit{LAT} catalog \cite{Fermi-LAT:2011iqa} as well as from \cite{Neronov:2012mx}. 
Each photon-like event detected at energies $E_\gamma>10$~GeV at high 
Galactic latitude $|b|>10^\circ$ of event class \texttt{ULTRACLEAN} and zenith angle $Z<100^\circ$
is matched against the list of AGN with known redshift. An event is considered to be associated with a
source, if its angular uncertainty ($r_{68}$ at 68~\%~c.l. derived from the instrumental response function P7V6\_ULTRACLEAN 
from the in-flight calibration) is
larger than the angular distance to the location of the AGN. 
 The resulting list of photons (see Fig.~\ref{Fig:photons}) 
contains 23(9) photons  with optical depth $\tau>1(2)$. Similar to the study carried out with VHE-spectra, we focus 
on the photons detected from sources at an optical depth $\tau>2$ (assuming the best-fit 
level of the extra-galactic background \cite{Ackermann:2012sza}).
 The two highest-energy photons exceeding the 
well-calibrated energy range of 500~GeV are excluded from the sample as well as four photons where the probability 
of association with the source is less than 90~\% even in the case of no absorption present\footnote{The probability 
is calculated using the \texttt{gtsrcprob} tool.}.
The final sample comprises three photons from GB6J1001+2911 ($\tau(E=308~\mathrm{GeV},z=0.558)=2.18$), 
S4 0218+35 ($\tau(E=179~\mathrm{GeV},z=0.944)=2.46$), Ton 599 ($\tau(E=301~\mathrm{GeV},z=0.725)=3.1$). For each photon
and source, the number of photons predicted from the source for a nominal absorption is calculated 
($=\mathcal{O}(10^{-3}$)) as well as the number of background events ($=\mathcal{O}(10^{-4})$). 
The predicted number of source photons is based upon a power-law extrapolation of the energy spectrum fit 
in the range from 1~GeV to the energy where absorption diminishes the expected flux by 1~\%. A power-law was
chosen even in the case of significant curvature of the energy spectrum. Given this choice, the predicted number of source
photons is an upper limit to the real value. \\
The resulting probability 
for detecting the three photons is combined using Fisher's method \cite{Fisher:1925} to be $P_\mathrm{pre-trial}=6.57\times10^{-6}(4.36~\sigma)$
and correcting for trials $P_\mathrm{post-trial}=1.17\times10^{-4}(3.68~\sigma)$, consistent with the result obtained from the VHE data.
Systematic effects include changing of the energy within the estimated $68~\%$ c.l. uncertainty 
($P_\mathrm{pre-trial}=3.34\times 10^{-5}$) and assuming a harder intrinsic spectrum ($P_\mathrm{pre-trial}=1.85\times10^{-5}$). In both
cases, the probabilities increase, but the significance remains larger than $3~\sigma$.

\section{Summary and discussion}
 We have extended our previous work to search for anomalous transparency of the
Universe to very-high energy (VHE) $\gamma$-rays to the low-energy regime
covered with the Fermi-\textit{LAT} instrument. We find three photons from
three sources with optical depth $\tau>2$. The combined probability to detect
these photons is (post-trial) $1.17\times10^{-4}$ corresponding to
$(3.68~\sigma)$. The on-going observation as well as improvements of
data-analysis will increase the sensitivity to search for deviations from the
expected (astrophysical) transparency for gamma-rays. Future observations
carried out with the next generation of ground-based air Cherenkov telescopes 
(Cherenkov telescope array: CTA \cite{Acharya:2013sxa}) 
will bridge the energy gap between the energy range covered with today's ground
based installation and space-based telescopes and therefore will be sensitive to confirm the indications for anomalous transparency.

\section*{Acknowledgements}
The authors thank the organizers of the Patras workshop. The work of MM 
has been funded through SFB 676 and the LEXI \textit{Connecting particles
with the cosmos}.


\begin{footnotesize}
\bibliography{main}
\bibliographystyle{plain}


%

\end{footnotesize}


\end{document}